\documentclass[11pt,letterpaper,floatfix]{article}
\pdfoutput=1

\usepackage{jcappub}
\usepackage{amssymb}
\usepackage{gensymb}
\usepackage{graphicx}
\usepackage{amsmath}
\usepackage{wasysym}
\usepackage{verbatim}

\begin{document}
\title{Constraints on WIMP and Sommerfeld-Enhanced Dark Matter
  Annihilation from HESS Observations of the Galactic Center}
\author[a]{Kevork N.\ Abazajian} \emailAdd{kevork@uci.edu}
\author[a,b]{J.\ Patrick Harding} \emailAdd{hard0923@umd.edu}
\affiliation[a]{Center for Cosmology, Department of Physics \& Astronomy,
  University of California,\\ Irvine, CA 92697}
\affiliation[b]{Maryland Center for Fundamental Physics, Department of
  Physics, University of Maryland,\\ College Park, MD 20742-4111}
\date{\today}

\abstract{ We examine the constraints on models of weakly interacting
  massive particle (WIMP) dark matter from the recent observations of
  the Galactic Center by the High Energy Spectroscopic System (HESS)
  telescope.  We analyze canonical WIMP annihilation into Standard
  Model particle final states, including $b\bar b$, $t\bar t$ and $W^+
  W^-$. The constraints on annihilation into $b\bar b$ is within an
  order of magnitude of the thermal cross section at $\sim$3 TeV,
  while the $\tau^+\tau^-$ channel is within a factor of $\sim$2 of
  thermal.  We also study constraints on Sommerfeld-enhanced dark
  matter annihilation models, and find that the gamma-ray
  observational constraints here rule out all of the parameter space
  consistent with dark matter annihilation interpretations of PAMELA
  and the Fermi-LAT $e^+e^-$ spectrum, in specific classes of models,
  and strongly constrains these interpretations in other classes.  The
  gamma-ray constraints we find  are more constraining on these models,
  in many cases, than current relic density, cosmic microwave
  background, halo shape and naturalness constraints.  }



\arxivnumber{arXiv:1110.6151}

\maketitle
\section{Introduction}
The existence of cosmological dark matter has been well-established by
observations of galaxy clusters, galaxy rotation curves, the cosmic
microwave background (CMB), and large-scale cosmological
structure. However the identity of the dark matter has remained a
fundamental unsolved problem in cosmology and particle physics for
nearly 80 years~\cite{Zwicky:1933gu}. Several particle candidates have
been proposed that could account for the dark matter (for a review,
see, e.g.~\cite{Feng:2010gw}). One well-motivated dark matter
candidate is a weakly-interacting massive particle (WIMP), which can
naturally produce a relic abundance at the observed dark matter
density. Thermal production of dark matter prefers a scale of the dark
matter cross-section at $\langle\sigma_{\rm A}v\rangle\approx 3\times
10^{-26}\rm\ cm^3\ s^{-1}$. This annihilation rate into Standard Model
particles results in energetic gamma-ray production through the
hadronization of quarks, bremsstrahlung of leptons, or directly into
two gammas through higher-order processes. This leads to the so-called
method of ``indirect detection'' of dark matter, constraining the dark
matter mass, annihilation cross-section, and annihilation spectrum
through a search for the Standard Model byproducts of WIMP
annihilation.

The High Energy Stereoscopic System (HESS) telescope has strong
sensitivity to high-energy gamma-rays such as those from high-mass
WIMPs~\cite{Abramowski:2011hc}. HESS consists of an
array of atmospheric \v{C}erenkov telescopes in Namibia designed to search
for high-energy gamma-rays, with 960 pixels per telescope at a
resolution of $0.16\degree$ per
pixel~\cite{Hinton:2004eu}. Specifically, HESS is sensitive to
gamma-ray energies from a few hundred GeV to a few tens of
TeV. Previously, studies of the Galactic Ridge for
$\lvert\ell\rvert<0.8\degree$ and $\lvert b\rvert<0.3\degree$ by
HESS~\cite{Aharonian:2006au} have been used to limit the dark matter
cross-section at high masses (0.5-30
TeV)~\cite{Regis:2008ij,Bertone:2008xr,Bell:2008vx,Mack:2008wu,Meade:2009iu,Bi:2009de,Abazajian:2010sq,Nekrassov:2011qk}. However,
an even more stringent constraint on the dark matter cross-section has
been shown to come from a new analysis of a region around the Galactic
Center (GC)~\cite{Abramowski:2011hc}. Here, we analyze in detail the
constraints arising from this observation.

The HESS GC analysis uses a reflected background technique to provide
a robust background region for the gamma-ray signal from the
GC. Because the background region is further from the GC than the
source region, it is expected that the dark matter signal should be
larger in the source region than the background region, for a dark
matter halo profile whose density peaks toward the GC. The HESS GC
analysis shows no excess gamma-ray signal in the source region over
the background region. Therefore, any dark matter annihilation signal
must produce few enough gamma-rays that the gamma-ray flux in the
source region is indistinguishable from the gamma-ray flux in the
background, at the current HESS sensitivity.

There have recently been stacked studies of the gamma-ray signals
coming from dwarf spheroidal galaxies by the Fermi-LAT which have been
used to constrain WIMP annihilation cross-sections and, in particular,
have excluded thermal low-mass WIMPs in the $b\bar{b}$ and
$\tau^+\tau^-$ annihilation channels below $\sim 30$
GeV~\cite{GeringerSameth:2011iw,collaboration:2011wa}. Here, we show
that the HESS GC analysis provides much tighter constraints than the
stacked Fermi-LAT dwarf spheroidals for dark matter masses above the
Fermi-LAT energy window.

There remains interest in the possibility of dark matter annihilation
as the source of the excess cosmic ray positron fraction at
$\sim10$-100 GeV observed by the PAMELA satellite, with $e^+e^-$ pairs
produced either directly or indirectly in a dark matter particle pair
annihilation
cascade~\cite{Adriani:2008zr,Cirelli:2008pk,Cholis:2008hb,Cholis:2008qq}. Additionally,
features in the higher-energy $10^2$ to $10^3$ GeV $e^+e^-$ spectrum seen
by the Fermi-LAT~\cite{Abdo:2009zk} are also consistent with the dark
matter annihilation interpretations of the lower energy positron
excess data~\cite{Cholis:2008wq,Bergstrom:2009fa,Meade:2009iu}. A recent study of
the $e^+e^-$ data from the Fermi-LAT is consistent with the positron
excess of the PAMELA satellite and shows the spectrum continuing to
rise up to at least $\sim 200$ GeV~\cite{Ackermann:2011rq}.

In order to achieve the dark matter annihilation rate required for
these $e^+e^-$ signals while remaining consistent with the expected
thermal production cross section, and to avoid an excess in
anti-proton observations (which is not seen), the annihilation rate
can be enhanced through a low-energy Sommerfeld-enhancement and
limited to leptonic modes with a light ($<1\rm\ GeV$) dark-force
carrying
particle~\cite{ArkaniHamed:2008qn,Pospelov:2008jd,Baumgart:2009tn,Katz:2009qq}. Such
an enhanced cross-section from a new force is in tension with detailed
calculations of the relic abundance of the dark matter, so that such a
candidate in many cases may not contribute to all of the dark
matter~\cite{Dent:2009bv,Zavala:2009mi,Feng:2009hw,Buckley:2009in,Feng:2010zp}.
Such candidates are also constrained by non-thermal distortions of the
CMB~\cite{Padmanabhan:2005es,Zavala:2009mi,Slatyer:2009yq,Hisano:2011dc}
and asphericity observed in dark matter
halos~\cite{Feng:2009hw,Buckley:2009in}. Cases of these models remain
viable given all such
constraints~\cite{Finkbeiner:2010sm,Slatyer:2011kg}.  There are also
constraints on these models from the observed diffuse gamma-ray and
X-ray backgrounds~\cite{Abazajian:2010zb,Zavala:2011tt}, observations
toward the GC by Fermi-LAT~\cite{Cirelli:2009dv}, as well as big-bang
nucleosynthesis~\cite{Hisano:2008ti,Hisano:2009rc}.

One model of dark matter with a light dark-force carrying particle and
Sommerfeld-enhanced cross-section is ``eXciting dark matter''
(XDM)~\cite{Finkbeiner:2007kk,ArkaniHamed:2008qn}. XDM was initially
proposed to explain the 511 keV gamma-ray signal from the GC (see
ref.~\cite{Teegarden:2004ct} for a discussion of these signals). In
XDM, there are two lowest-energy dark states $\chi$ and $\chi^*$ which
have masses differing by only a few MeV, with a light gauge bosons $\phi$
mediating excitations from $\chi$ to $\chi^*$. The exchange of many
gauge bosons leads to a Sommerfeld-enhanced cross-section much larger than
that of a thermal relic. The annihilation
$\chi\chi\rightarrow\phi\phi$ followed by the decay of $\phi$ into
leptons leads to an excess of high-energy electrons and positrons in
the GC. This scenario has also been considered to explain the excess
in local positrons seen by the PAMELA and Fermi-LAT satellites. Most
recently, ref.~\cite{Finkbeiner:2010sm} has interpreted the XDM
explanation for the PAMELA and Fermi-LAT excesses including constraints from
the thermal relic density, CMB, self-interaction bounds, and
naturalness bounds. Below, we compare the gamma-ray constraints from
the HESS GC analysis to these other XDM limits.

The sensitivity of the HESS GC observation to dark matter
annihilation, as with all Galactic Center observations, depends on the
nature of the dark matter density profile.  Specifically, for the HESS
GC background subtraction method, there must be a higher dark matter
density within the inner $\lesssim$150~pc, in the signal region, than
the background subtraction region between approximately 150~pc and
450~pc.  As pointed out also in the HESS Collaboration
analysis~\cite{Abramowski:2011hc}, if there exists a constant-density
core within the inner $\sim$450~pc of the Milky Way, no limits on dark
matter annihilation can be derived from the HESS GC observation since
the background subtraction would also remove any equivalent signal.

There is significant debate in the literature as to the nature of the
inner dark matter profile of a galaxy such as the Milky Way. Numerical
simulations are employed in attempts to accurately determine the inner
dark matter density profile.  The canonically-adopted dark matter halo
density profile for the case of cold dark matter is the $r^{-1}$
inner-radius scaling Navarro-Frenk-White (NFW) profile
~\cite{Navarro:1996gj}.  The highest spatial resolution simulations of
Milky-Way-type halo formation are pure dark matter halo simulations:
Via Lactea II~\cite{Diemand:2008in}, GHALO~\cite{Stadel:2008pn} and
AQUARIUS (A-1)~\cite{Navarro:2008kc} (which have gravitational
softening lengths of 40, 61 and 20~pc, respectively), which find a
peaked density profile down to $\sim$100~pc, with a
logarithmically-changing slope that is sometimes dubbed an ``Einasto''
profile.  The vast majority of studies that include baryons in
addition to dark matter have found that baryonic effects concentrate
and steepen the central dark matter distribution due to adiabatic
contraction~\cite{Blumenthal:1986,Gnedin:2004cx}, including recent
high-resolution simulations with gas cooling, star formation, and
stellar feedback
processes~\cite{Pedrosa:2009rw,Tissera:2009cm,Guedes:2011ux,Gnedin:2011uj}.
Importantly, ref.~\cite{Gnedin:2011uj} does an extensive error
analysis of their numerical results.  In contrast, some studies'
simulations have claimed that baryonic effects may have the opposite
effect, reducing the dark matter density in the central region via
dark matter expansion from stellar and gas feedback outflows, and
producing flat or nearly-flat density cores at up to 2-3~kpc in size
for an approximately Milky Way size halo, fit by cored isothermal or
Burkert
profiles~\cite{RomanoDiaz:2008wz,RomanoDiaz:2009yq,Maccio':2011eh}.
It has been shown in ref.~\cite{Zhan:2005eg} that the S2 or Plummer
force softening must be a factor of $\approx$5 times smaller than the
scale of interest for the inner profile of dark matter halos in order
to achieve greater than 5\% accuracy in radial accelerations of
particles, with an ideal time-step algorithm choice.  Poor force
resolution has been shown to generally lead to artificially lower
central densities~\cite{Power:2002sw}.  The gravitational softening
length used in the simulations in ref.~\cite{RomanoDiaz:2009yq} is
0.5~kpc, so it is questionable to draw conclusions at the claimed
$\sim$kpc core scale.  In the recent work of
ref.~\cite{Maccio':2011eh}, the gravitational softening length is
0.3125~kpc~\cite{Stinson:2010xe}, and therefore conclusions of the
inner 1-2~kpc are also difficult to make with confidence.
Furthermore, ref.~\cite{Maccio':2011eh} finds evidence for a core to
be produced only in the more extreme feedback High Feedback Run, while
the Low Feedback Run found adiabatic contraction that steepened the
dark matter profile.

In summary, work indicating the presence of dark matter density cores
in numerical simulations are at the edge of their resolution limits.
It is important to consider that if it becomes firmly established from
numerical simulations that there is necessarily a constant-density
core in the Milky Way at the $\gtrsim 450\rm\ pc$ scale, then the HESS
GC limits presented here are not applicable due to the observation's
background subtraction method.  This was also noted by the HESS
collaboration work~\cite{Abramowski:2011hc}.  However, at this time
robust numerical simulations predominantly indicate contraction and
steepening of the central density profile due to baryonic effects, and
the adoption of the non-contracted NFW or Einasto profile here is
conservative relative to the steeper profiles.

Below we show that, in the case of a non-adiabatically-contracted NFW
or Einasto dark matter halo profile, the HESS GC observation provides
a strong limit on the cross-section of high-mass WIMPs' annihilation
into several Standard Model channels. Furthermore, we show how the
dark matter annihilation interpretation of the PAMELA excess and
Fermi-LAT $e^+e^-$ feature signals is excluded at above $95\%$ CL in
many cases.  Dark matter interpretations of these signals are in
tension with the HESS GC observations for two-body standard model
particle final states and gauge-boson mediated four-lepton final states,
when also including the constraints for Fermi-LAT observations toward
local dwarf galaxies. Importantly, the HESS limits presented here for
XDM are more constraining than the thermal relic density, CMB,
self-interaction bounds, and naturalness bounds.
\section{Data Analysis}
The data used in this paper comes from the HESS collaboration analysis
of the GC from ref.~\cite{Abramowski:2011hc}. The events analyzed are
those from 112 hours of live time from the HESS very-high energy
gamma-ray instrument with zenith angles smaller than $30\degree$ which
were within the central $4\degree$ of the HESS
field-of-view. Contamination of the dark matter signal due to the
Galactic plane is excluded by masking the regions with Galactic
latitude $\lvert b\rvert<0.3\degree$ and an additional mask within
$0.3\degree$ of the extended source HESS J1745\textendash303
($(b,\ell)=(-0.6\degree,358.71\degree)$)~\cite{Aharonian:2005kn}.

The source region is defined by $0.02\degree\times 0.02\degree$ pixels
that lie within $1\degree$ of the GC, do not lie within the mask, and
have a well-defined background. The background region is determined by
choosing a telescope pointing position within $1.5\degree$ of the GC
and rotating each pixel in the source region (with masked pixels
removed) by $90\degree$, $180\degree$, and $270\degree$ about the
telescope pointing position. Any of these pixels which is further than
$1\degree$ from the GC and does not lie within the mask is considered
background. Pixels within the inner $1\degree$ for which there are no
corresponding background pixels are excluded from the analysis. This
process was then repeated for multiple telescope pointing positions,
calculating the differential flux in both the background and source
regions. For additional details on the HESS GC analysis, see
ref.~\cite{Abramowski:2011hc}.
\section{Dark Matter Annihilation Limits from the HESS GC Observations}
\subsection{Gamma-Ray Emission from Annihilating Dark Matter}
A robust calculation of the expected final state radiation from dark
matter annihilation requires accurate quantification of the dark
matter source as well as the products in the final state gamma-ray
radiation chain. The differential flux per solid angle for a dark
matter candidate with cross-section $\langle\sigma_{\rm A}v\rangle$
over a solid angle $\Delta\Omega$ is
\begin{equation}
\frac{dF}{dE}=\frac{\langle\sigma_{\rm A}v\rangle}{2}\frac{J_{\Delta\Omega}}{J_0}\frac{1}{4\pi M_\chi^2}\frac{dN_\gamma}{dE}\enspace,
\end{equation}
where $dN_\gamma/dE$ is the gamma-ray spectrum per annihilation and
$M_\chi$ is the dark matter particle mass. The integrated mass density
squared along line-of-sight $x$, averaged over the solid angle of the
observation region $\Delta\Omega$ is defined as
\begin{equation}
J_{\Delta\Omega}=\frac{J_0}{\Delta\Omega}\int_{\Delta\Omega}d\,\Omega\int d\,x\ \rho^2(r_{\rm gal}(b,\ell,x))\enspace,
\end{equation}
where distance from the GC is given by
\begin{equation}
r_{\rm gal}(b,\ell,x)=\sqrt{R_{\odot}^2-2xR_{\odot}\cos(\ell)\cos(b)+x^2}\enspace.
\end{equation}
A normalization constant $J_0\equiv 1/\left[8.5\ {\rm kpc}(0.3\ {\rm
    GeV\ cm^{-3}})^2\right]$ is chosen to make $J$ dimensionless, but
the final flux calculation is independent of the choice of $J_0$. The
coordinates $b$ and $\ell$ are the Galactic latitude and longitude,
respectively. Following the HESS Collaboration, the dark matter
profiles $\rho(r)$ we choose are the non-adiabatically-contracted
Einasto and NFW models of ref.~\cite{Pieri:2009je} with the local dark
matter density $\rho_{\odot}\equiv 0.389\rm\ GeV\ cm^{-3}$ ($0.385\rm\
GeV\ cm^{-3}$) for an NFW (Einasto)
profile~\cite{Catena:2009mf}. Since there were several telescope
pointing positions not given explicitly in
ref.~\cite{Abramowski:2011hc}, we cannot and did not independently
calculate $\bar{J}\,_{\Delta\Omega}$ but adopt those of the HESS
Collaboration.  Importantly, as discussed in the introduction and in
the HESS Collaboration study, in the case of a cored-isothermal or
Burkert profile with a constant-density core that extends at or beyond
$\sim$450~pc, then the background subtraction region would have an
identical annihilation signal as the signal region, and no constraint
can be placed on dark matter annihilation by this method.  We have
checked that the $J$-values for the single pointing shown in figure 2 of
ref.~\cite{Abramowski:2011hc} are approximately their
$\bar{J}$'s. Averaging over telescope pointing positions, the HESS
analysis calculated the dark matter $J$-values for the source
($\bar{J}\,\rm_s$) and background ($\bar{J}\,\rm_b$) regions using the
NFW and Einasto profiles~\cite{Abramowski:2011hc}:
\begin{align}
\bar{J}\,_{\rm s}^{\rm NFW}&=1604\label{J1}\\
\bar{J}\,_{\rm b}^{\rm NFW}&=697\\
\bar{J}\,_{\rm s}^{\rm Einasto}&=3142\\
\bar{J}\,_{\rm b}^{\rm Einasto}&=1535\label{J4}\enspace.
\end{align}
Note that toward the GC we adopt zero astrophysical substructure
boost.  However, there may be a substructure contribution to the flux
of order 4\% to 0.04\% in the inner
$\sim$1$^\circ$ view of the GC \cite{Kuhlen:2008aw}.
\subsection{Calculation of Dark Matter Spectra}
To calculate the photon spectrum for a particular WIMP annihilation
channel, we use {\sc pythia 6.4} to simulate the photon radiation of
charged particles as well as decays of particles such as the
$\pi^0$~\cite{Sjostrand:2006za}. Specifically, we run {\sc pythia} to
simulate an $e^+e^-$ collision at a center of mass energy of $2M_\chi$
through a $Z'$ to a final state that corresponds to the annihilation
products of the dark matter. For WIMP annihilation to final states
through light gauge bosons (such as the $4e$ and $4\mu$ channels) we have
the $Z'$ decay into two scalar ($H_0$) states, each of which
annihilates into two leptons. For the XDM case, we employ a light
gauge boson which has
branching fraction to $e^+e^-$, $\mu^+\mu^-$, and $\pi^+\pi^-$ set by
the particular XDM model. For the models with light gauge bosons, photons
can only be radiated at fairly low energy in the gauge boson rest
frame. Therefore, in the center-of-mass frame of the dark matter
annihilation, the average number of hard photons is significantly
reduced in comparison to direct annihilation to a two-body standard
model final state.

We turn off initial state radiation such that all photons only come
from the radiation or decay of the dark matter annihilation
products. We turn on the decays of particles which are not decayed
with the default {\sc pythia} settings, namely muons, charged pions, and
charged kaons. Additionally, we turn on the muon decay channel
$\mu^-\rightarrow e^-\nu_\mu\bar{\nu}_e\gamma$, with the standard
branching fraction of $0.014$~\cite{Nakamura:2010zzi}. Using a large
sample of events for each final state and each value of $M_\chi$, the
number of photons in the final state in a given logarithmic energy bin
is counted and averaged over the number of events, yielding the
average number of photons in that energy bin per annihilation event.

For WIMP annihilation to final states through extremely light gauge bosons
($m_\phi\lesssim0.5\rm\ GeV$), the {\sc pythia} calculation becomes impractical
due to low-energy cutoffs for radiation processes. For such
annihilation channels (with $m_\phi=0.25\rm\ GeV$ and
$m_\phi=0.35\rm\ GeV$) we instead use the analytic formulae given in
Appendix A of ref.~\cite{Mardon:2009rc}. For the $4e$ state, all
photons come from radiation off the final electrons and positrons. For
the $4\mu$ state, photons come from radiation off the final electrons
and positrons as well as the radiation from the muons before they
decay. We have verified the analytic spectra model by comparison to
{\sc pythia} spectra for heavier gauge bosons.
\subsection{Limits on the Dark Matter Annihilation Cross-Section}
\begin{figure*}[ht]
\begin{center}$
\begin{array}{c}
\includegraphics[width=0.68\columnwidth]{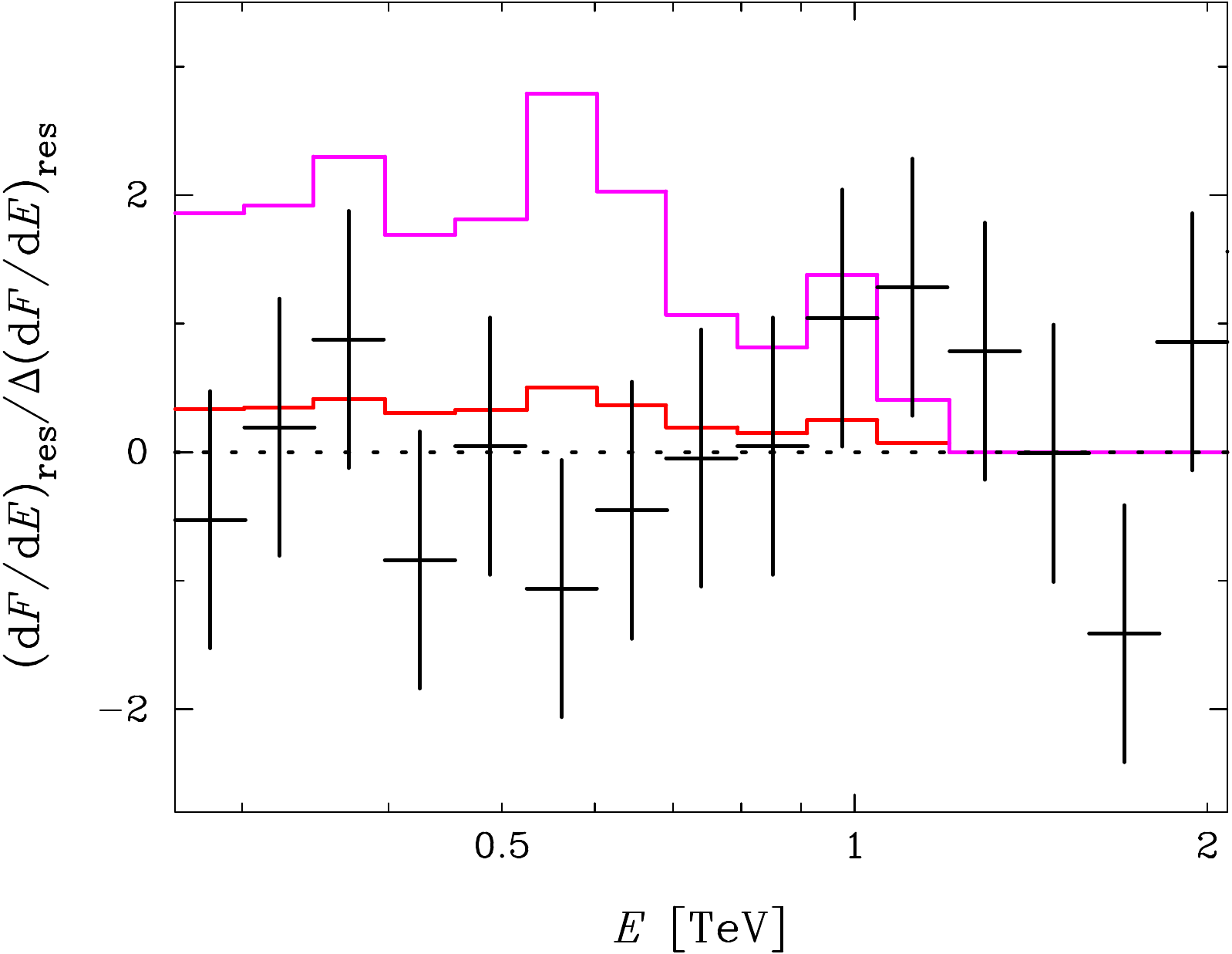}
\end{array}$
\end{center}
\caption{\small Shown are the data from the HESS GC observation, for
  which the signal and background are consistent with no difference in
  flux at the level of $\chi^2/{\rm DOF}=0.75$. For comparison, two
  possible dark matter signals are shown as well. Both dark matter
  signals are for a 1.2 TeV WIMP annihilating via 0.25 GeV gauge
  bosons into two $e^+e^-$ pairs, for an NFW halo profile. The red
  (lower) histogram is the signal expected for a dark matter
  cross-section $\langle\sigma_{\rm
    A}v\rangle=9\times10^{-25}\rm\ cm^3\ s^{-1}$, which has a total
  $\Delta \chi^2 = 2.79$. The magenta (upper) histogram is the signal
  expected for a dark matter cross-section $\langle\sigma_{\rm
    A}v\rangle=5\times10^{-24}\rm\ cm^3\ s^{-1}$ and mass $M_\chi =
  1.2\rm\ TeV$, with a $\Delta \chi^2 = 41.2$. These signals
  correspond with the upper end of the red Fermi-LAT plus PAMELA
  region, and lower end of the purple rectangle in
  figure~\ref{4leptons}(a), in the case of assuming a subdominant
  substructure contribution, and are excluded at greater than $95\%$
  CL. Note that the HESS GC observation extends to higher energies
  than shown here.\label{signalvsnoise}}
\end{figure*}
Figure 3 of ref.~\cite{Abramowski:2011hc} shows the observed source
and background fluxes in 35 energy bins from 0.28 to 31 TeV. The two
regions are consistent with each other, with zero difference in flux
having a $\chi^2/{\rm DOF}=0.75$. Therefore, any dark matter signal
must be small enough that the source region does not have appreciably
greater gamma-ray flux, within errors, than the background
region. Figure~\ref{signalvsnoise} illustrates the difference between
the lack of a dark matter signal from HESS and two representative dark
matter signals. Using the dark matter $J$-values from
eqs.~\ref{J1}\textendash\ref{J4}, we have derived $95\%$ one-sided
confidence-level (CL) (corresponding with a $90\%$ two-sided CL)
constraints on the dark matter cross-section as a function of mass for
several key annihilation channels, using the total $\chi^2$ in all bins.

\begin{figure*}[h]
\begin{center}$
\begin{array}{c}
\includegraphics[width=0.67\columnwidth]{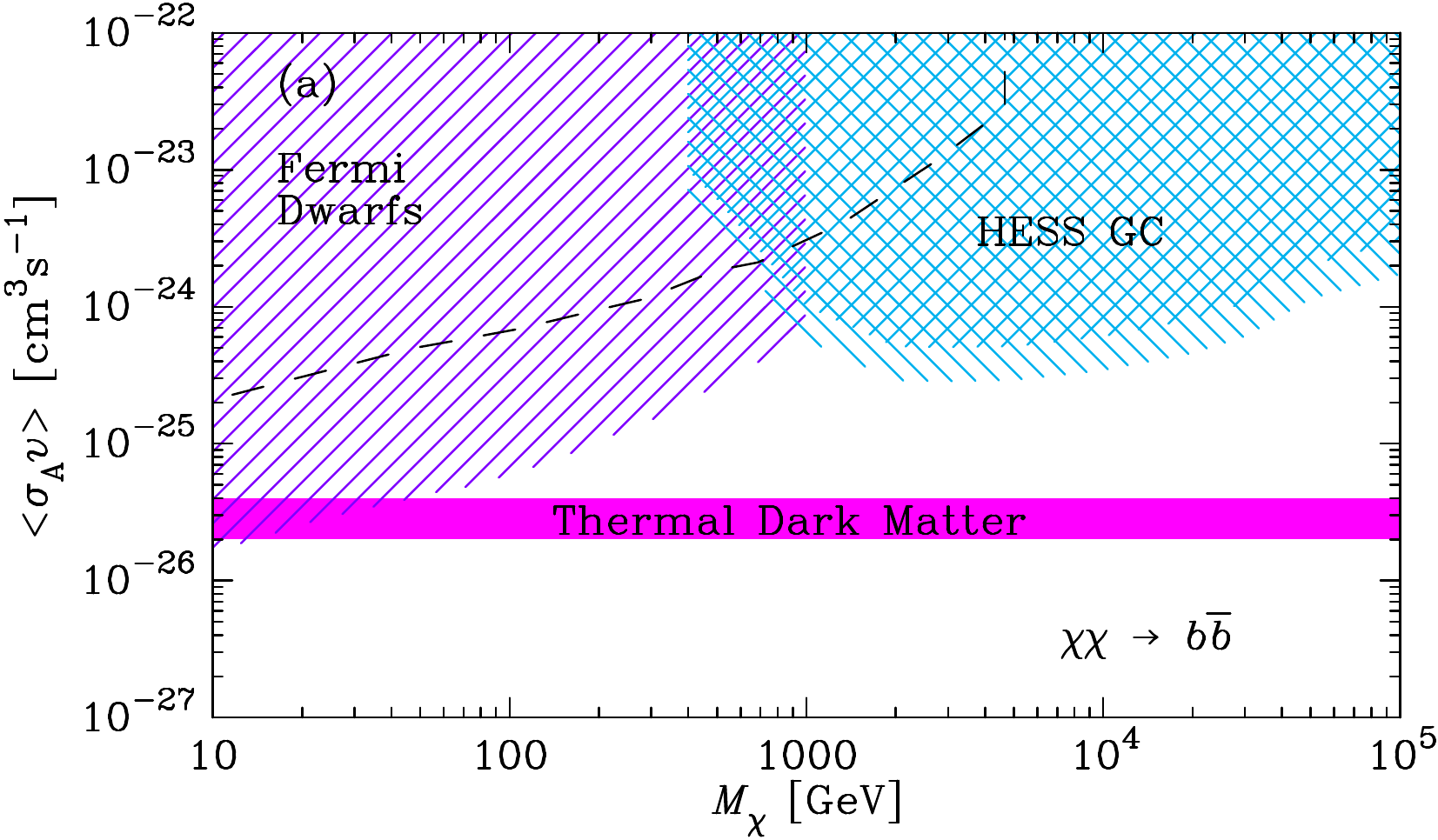} \\
\includegraphics[width=0.67\columnwidth]{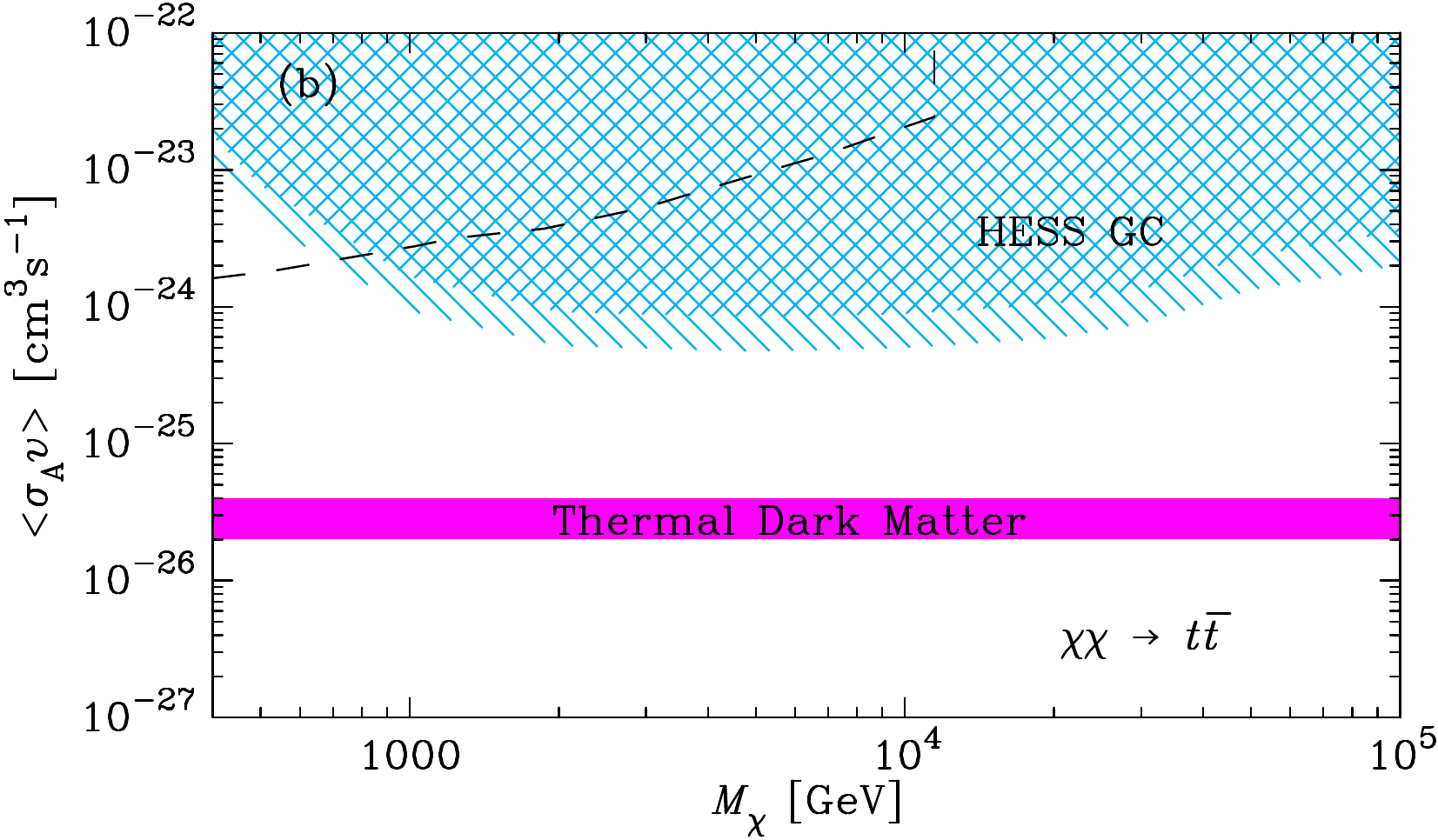} \\
\includegraphics[width=0.67\columnwidth]{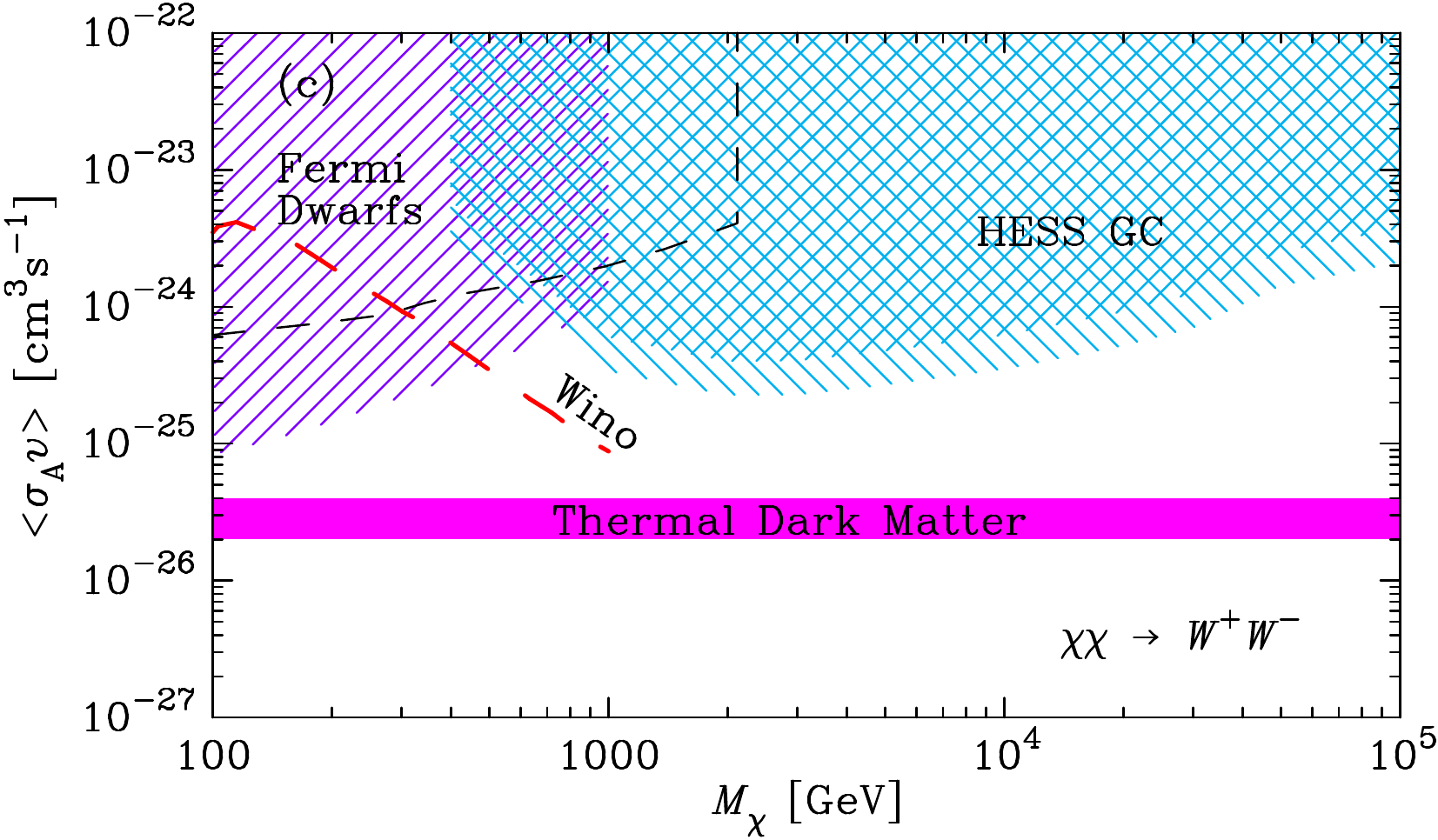}
\end{array}$
\end{center}
\caption{\small Shown are the constraints on dark matter in three
  canonical annihilation channels: (a) $b\bar{b}$; (b) $t\bar{t}$; (c)
  $W^+W^-$. The regions are labeled according to their constraining
  observations as described in the text: ``HESS GC'' are the $95\%$ CL
  limits from the HESS analysis of the GC. The double hatched region
  is constrained for both the Einasto and NFW halo models, and the
  single hatched region is constrained for only the Einasto halo
  model. The regions labeled ``Fermi Dwarfs'' are the $95\%$ CL limits
  from the Fermi-LAT collaboration analysis of dwarf spheroidals. In
  the $W^+W^-$ channel, panel (c), the mass for a non-thermal
  wino-like neutralino is shown as a thick-dashed red
  line~\cite{Grajek:2008jb}. For comparison, we plot the $3\sigma$
  limits from ref.~\cite{Cirelli:2009dv} for their analysis of the
  Fermi-LAT observation of the $3^\circ\times 3^\circ$ region around
  the Galactic Center as dashed (black) lines in all panels for the
  respective channels.\label{hadronic}}
\end{figure*}

\begin{figure*}[ht]
\begin{center}$
\begin{array}{cc}
\includegraphics[width=0.95\columnwidth]{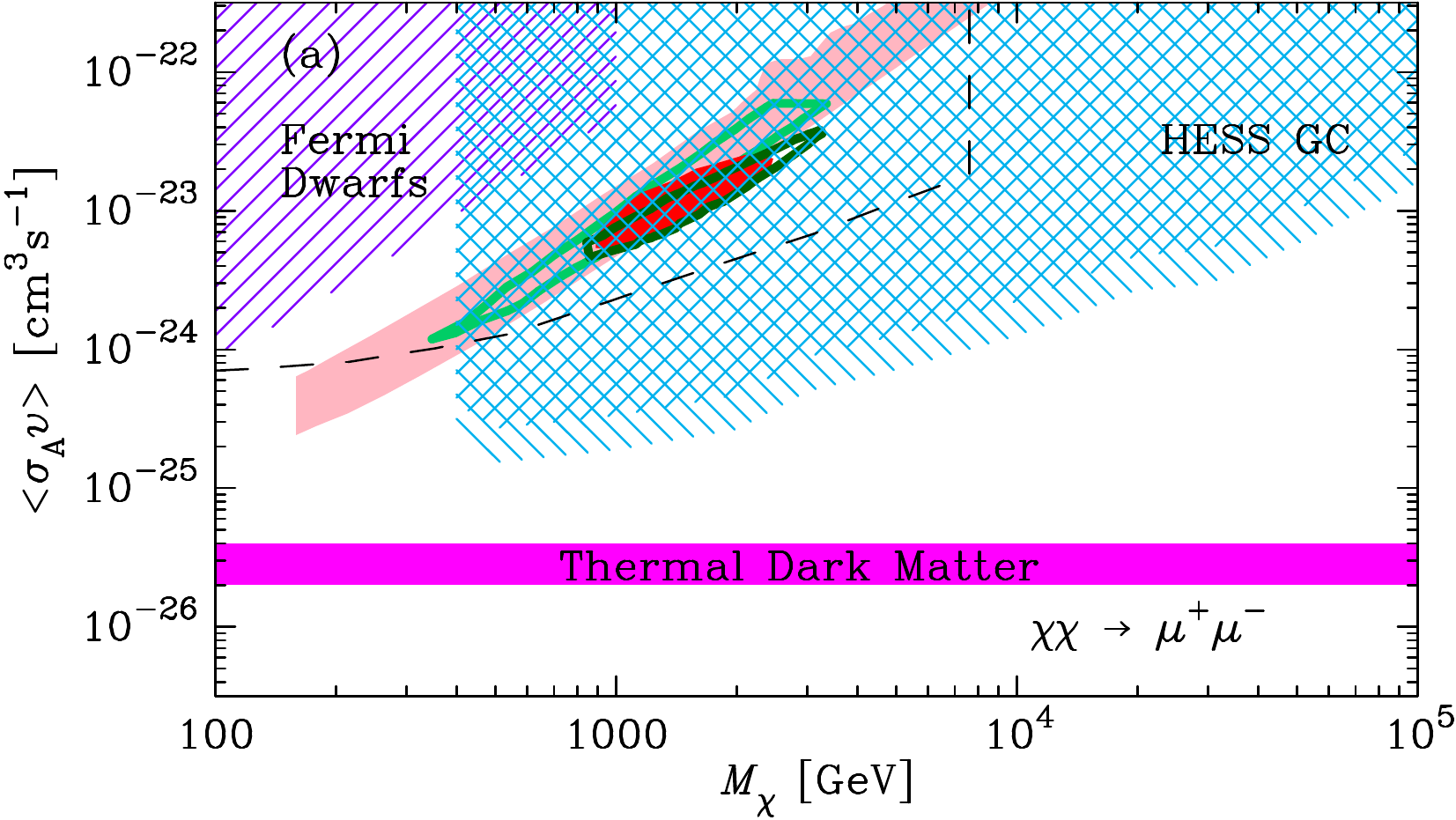} \\
\includegraphics[width=0.95\columnwidth]{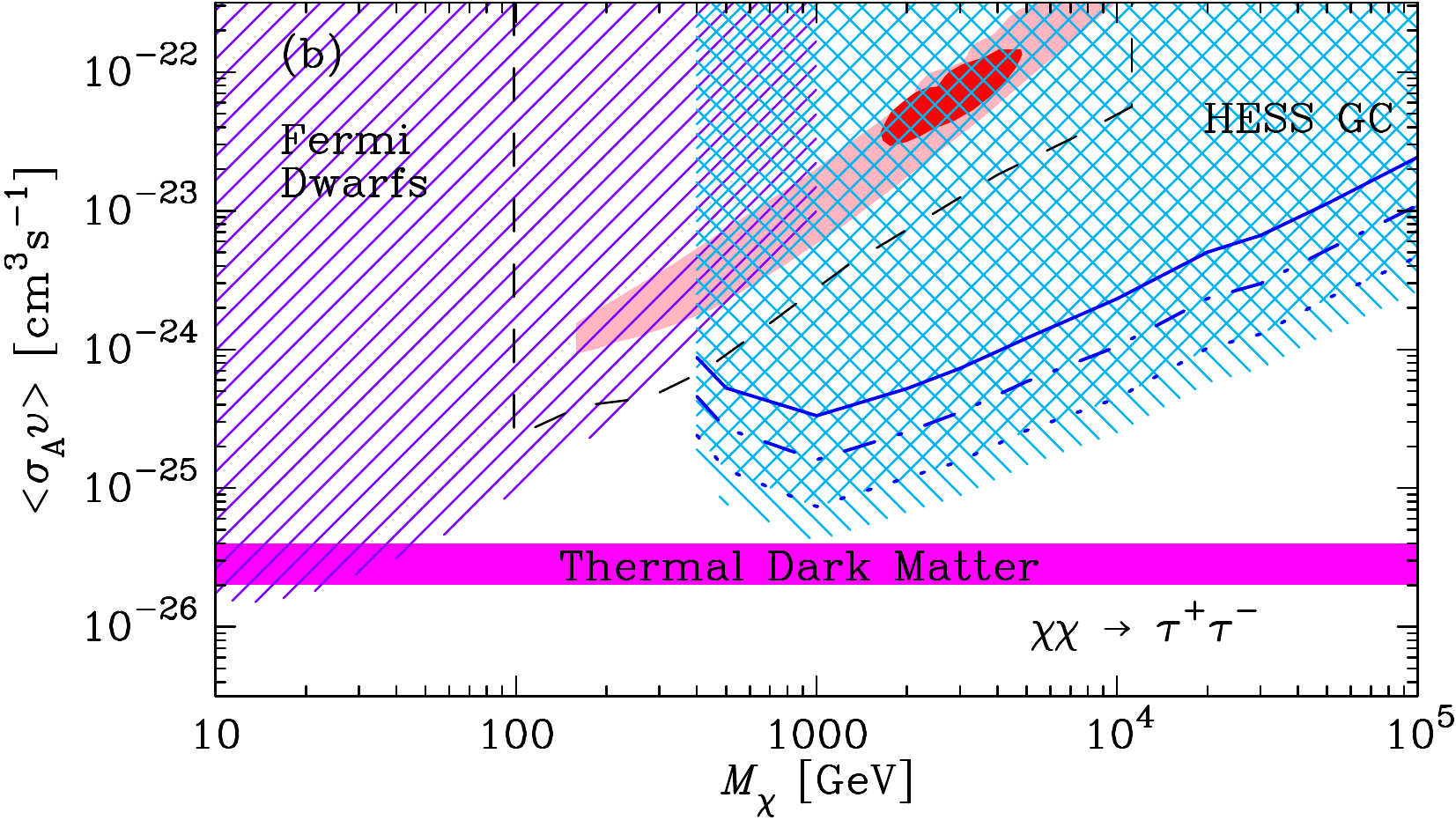}
\end{array}$
\end{center}
\caption{\small Shown are the constraints on dark matter in two
  leptonic annihilation channels: (a) $\mu^+\mu^-$; (b)
  $\tau^+\tau^-$. The regions are labeled according to their
  constraining observations as described in the text: ``HESS GC'' are
  the $95\%$ CL limits from the HESS analysis of the GC. The double
  hatched region is constrained for both the Einasto and NFW halo
  models, and the single hatched region is constrained for only the
  Einasto halo model. The regions labeled ``Fermi Dwarfs'' are the
  $95\%$ CL limits from the Fermi-LAT collaboration analysis of dwarf
  spheroidals. The light pink shaded region is consistent with a dark
  matter interpretation of the PAMELA signal and the dark red shaded
  region is that from the Fermi-LAT $e^+e^-$ feature from
  ref.~\cite{Meade:2009iu}. In the $\mu^+\mu^-$ channel, panel (a),
  the light green outlined region is consistent with a dark matter
  interpretation of the PAMELA signal and the dark green outlined
  region with that of the Fermi-LAT $e^+e^-$ feature from
  ref.~\cite{Bergstrom:2009fa}. PAMELA regions below
  $M_\chi=160\rm\ GeV$ are ruled out by the rise in the positron
  fraction seen by the Fermi-LAT~\cite{Ackermann:2011rq}. In panel
  (b), to illustrate the strength of the HESS GC limits, we show for
  the NFW profile the 95\%, 99.7\% and 99.9999\% CL limits in dotted,
  dot-dashed and solid lines, respectively.  For comparison, we plot
  the $3\sigma$ limits from ref.~\cite{Cirelli:2009dv} for their
  analysis of prompt and inverse-Compton emission in the Fermi-LAT
  observation of the $3^\circ\times 3^\circ$ region around the
  Galactic Center as dashed (black) lines in both panels for the
  respective channels.\label{leptonic}}
\end{figure*}

In the figures, the light blue cross-hatched region is excluded at
$95\%$ CL for both the NFW and Einasto dark matter profiles. The
singly-hatched light blue regions are excluded at $95\%$ CL for the
Einasto profile but not for the more conservative NFW profile. The
purple regions are the $95\%$ CL limits from a combined analysis of
ten dwarf spheroidal galaxies from the Fermi-LAT
Collaboration~\cite{collaboration:2011wa}. In particular, note that in
figure~\ref{hadronic}(a) and figure~\ref{leptonic}(b) we have extended
the mass of the dark matter down to 10 GeV in order to show the
exclusion of a standard thermal relic below 27 GeV (37 GeV) for the
$b\bar{b}$ ($\tau^+\tau^-$) annihilation channel by the Fermi-LAT
stacked analysis of dwarf galaxies~\cite{collaboration:2011wa}. The
dark pink regions represent the annihilation cross-section for a
thermal relic,
$\langle\sigma_{\rm A}v\rangle\approx3\times10^{-26}\rm\ cm^3\ s^{-1}$. The
dark matter constraints for the standard WIMP annihilation channels
$b\bar{b}$, $t\bar{t}$, and $W^+W^-$ are shown in
figure~\ref{hadronic}. We show the $t\bar{t}$ channel since this
should dominate for high mass dark matter ($M_\chi \gtrsim
200\rm\ GeV$). Our $b\bar{b}$ constraints are consistent with the
quark channel limits from HESS~\cite{Abramowski:2011hc}. As a
comparison to the thermal cross-section, figure~\ref{hadronic}(c) also
includes the expected cross-section for a non-thermal wino-like
neutralino (the dashed red line)~\cite{Grajek:2008jb}.

\begin{figure*}[ht]
\begin{center}$
\begin{array}{c}
\includegraphics[width=0.9\columnwidth]{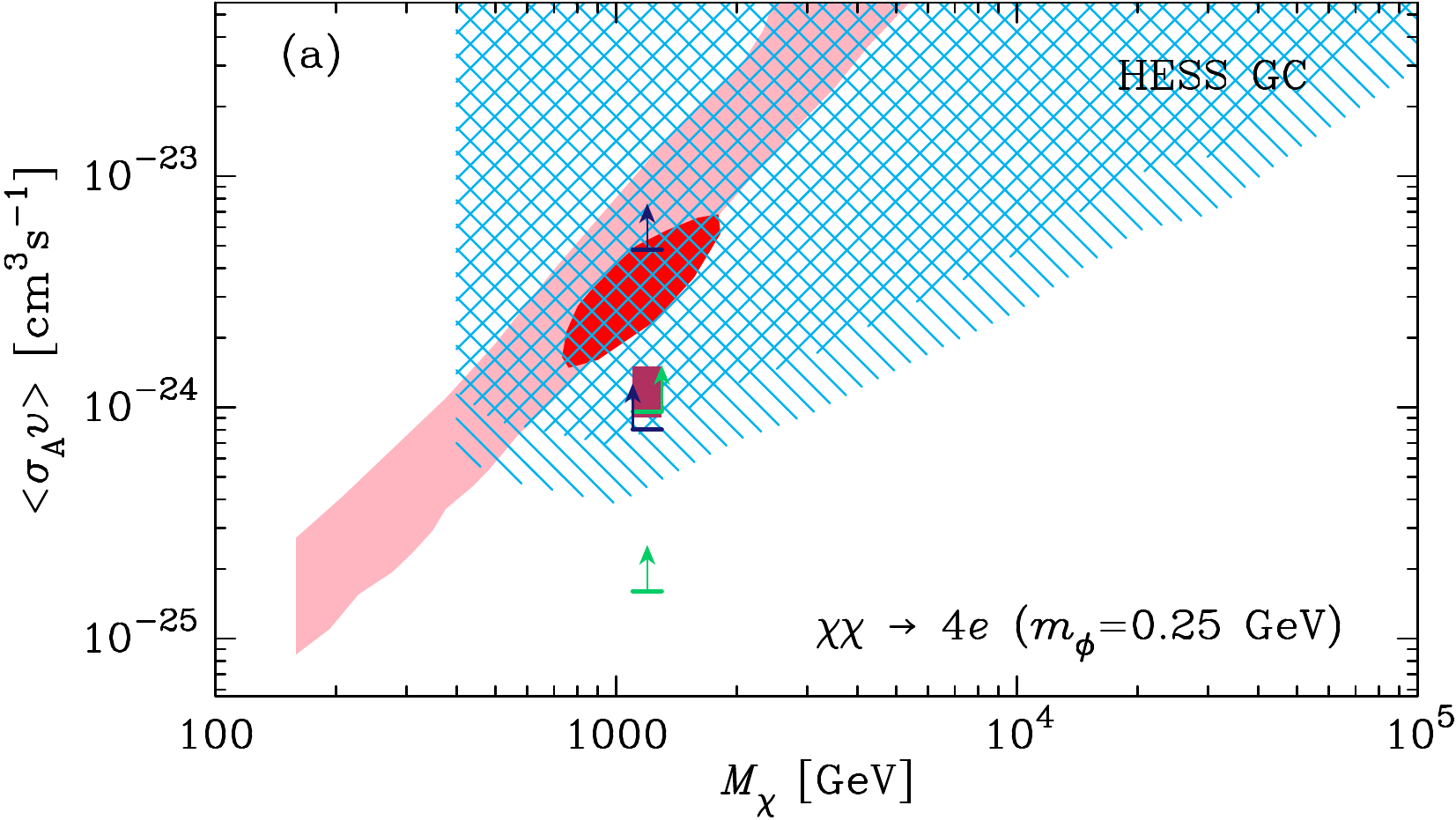} \\
\includegraphics[width=0.9\columnwidth]{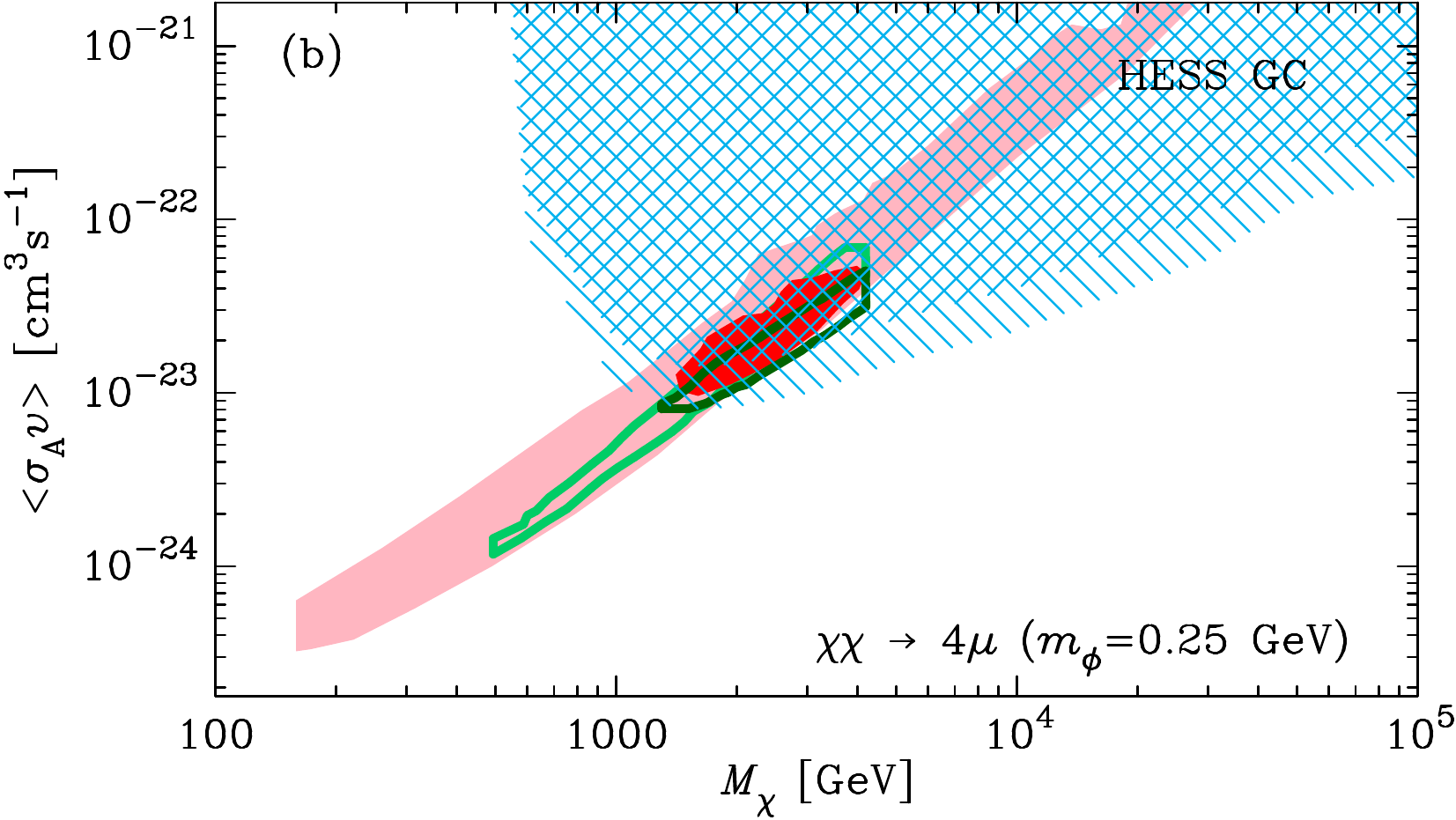}
\end{array}$
\end{center}
\caption{\small Shown are the constraints on dark matter in two
  annihilation channels: (a) annihilation into two $e^+e^-$ pairs via
  two intermediate 0.25 GeV gauge bosons $\phi$ (note that this case
  requires an {\it ad hoc} requirement of decay into electrons and not
  muons); (b) annihilation into two $\mu^+$ and two $\mu^-$ via two
  intermediate 0.25 GeV gauge bosons $\phi$. The regions are labeled
  according to their constraining observations as described in the
  text: ``HESS GC'' are the $95\%$ CL limits from the HESS analysis of
  the GC. The double hatched region is constrained for both the
  Einasto and NFW halo models, and the single hatched region is
  constrained for only the Einasto halo model. The light pink shaded
  region is consistent with a dark matter interpretation of the PAMELA
  signal and the dark red shaded region is that of the Fermi-LAT
  $e^+e^-$ feature from ref.~\cite{Meade:2009iu}. In the $4e$ channel,
  panel (a), the purple rectangle demonstrates the range of
  Sommerfeld-enhanced cross-sections consistent with constraints from
  thermal relic density, the CMB, self-interaction bounds, and
  naturalness~\cite{Slatyer:2011kg}. The green (blue) bars and arrows
  show the HESS GC limits for the two cases of velocity dispersions of
  $v\rightarrow 0$ ($v\sim\rm 150\ km\ s^{-1}$), respectively, with
  the upper and lower bars for each color representing the two local
  substructure boost limits, as described in the text.  In the $4\mu$
  channel, panel (b), the light green outlined region is consistent
  with a dark matter interpretation of the PAMELA signal and the dark
  green outlined region is that for Fermi-LAT $e^+e^-$ feature from
  ref.~\cite{Bergstrom:2009fa}. PAMELA regions below $M_\chi\approx
  160\rm\ GeV$ are ruled out by the rise in the positron fraction seen
  by the Fermi-LAT~\cite{Ackermann:2011rq}.\label{4leptons}}
\end{figure*}

Figure~\ref{leptonic} shows the dark matter limits for dark matter
annihilating directly into leptons, whereas in figure~\ref{4leptons},
the dark matter annihilates into two gauge bosons $\phi$ of mass
$m_\phi=0.25\rm\ GeV$ which then decay into leptons. (Note that the
4$e$ case requires an {\it ad hoc} requirement of gauge boson decay
into electrons and not muons. Neither the constraints nor
signals would be significantly different in the case of, e.g., $m_\phi
= 0.2\rm\ GeV$, where annihilation to $e^+ e^-$ would be energetically
required.) Heavy dark matter masses annihilating primarily into
leptons are particularly interesting in the context of the PAMELA
positron excess~\cite{Adriani:2008zr} and the $e^+e^-$ feature seen by
the Fermi-LAT~\cite{Abdo:2009zk}. Such leptonic annihilation channels
with a cross-section much larger than that of a thermal relic have
been studied as the source of these
anomalies~\cite{Meade:2009iu,Bergstrom:2009fa}. In
figures~\ref{leptonic} and~\ref{4leptons} we have included one dark
matter annihilation interpretation of the PAMELA excess in light pink
and the analogous interpretation of the Fermi-LAT feature in
red~\cite{Meade:2009iu}. Recent analysis of the positron fraction by
the Fermi-LAT Collaboration consistent with the PAMELA excess
continues to rise up to energies of 180 GeV bin center energy, ruling
out PAMELA regions below $M_\chi\approx 160\rm\ GeV$ from being
consistent with dark matter annihilation~\cite{Ackermann:2011rq}. In
figures~\ref{leptonic}(a) and~\ref{4leptons}(b) we include fits to the
interpretation of the PAMELA excess in light green outline and the
Fermi-LAT $e^+e^-$ feature in dark green outline from
ref.~\cite{Bergstrom:2009fa}.  In panel \ref{4leptons}(b), for the NFW
profile case, we exemplify the strength of the limit by plotting the
95\%, 99.7\%, and 99.9999\% CL limits as dashed, dot-dashed and solid
lines, respectively.

\begin{figure*}[ht]
\begin{center}$
\begin{array}{c}
\includegraphics[width=0.68\columnwidth]{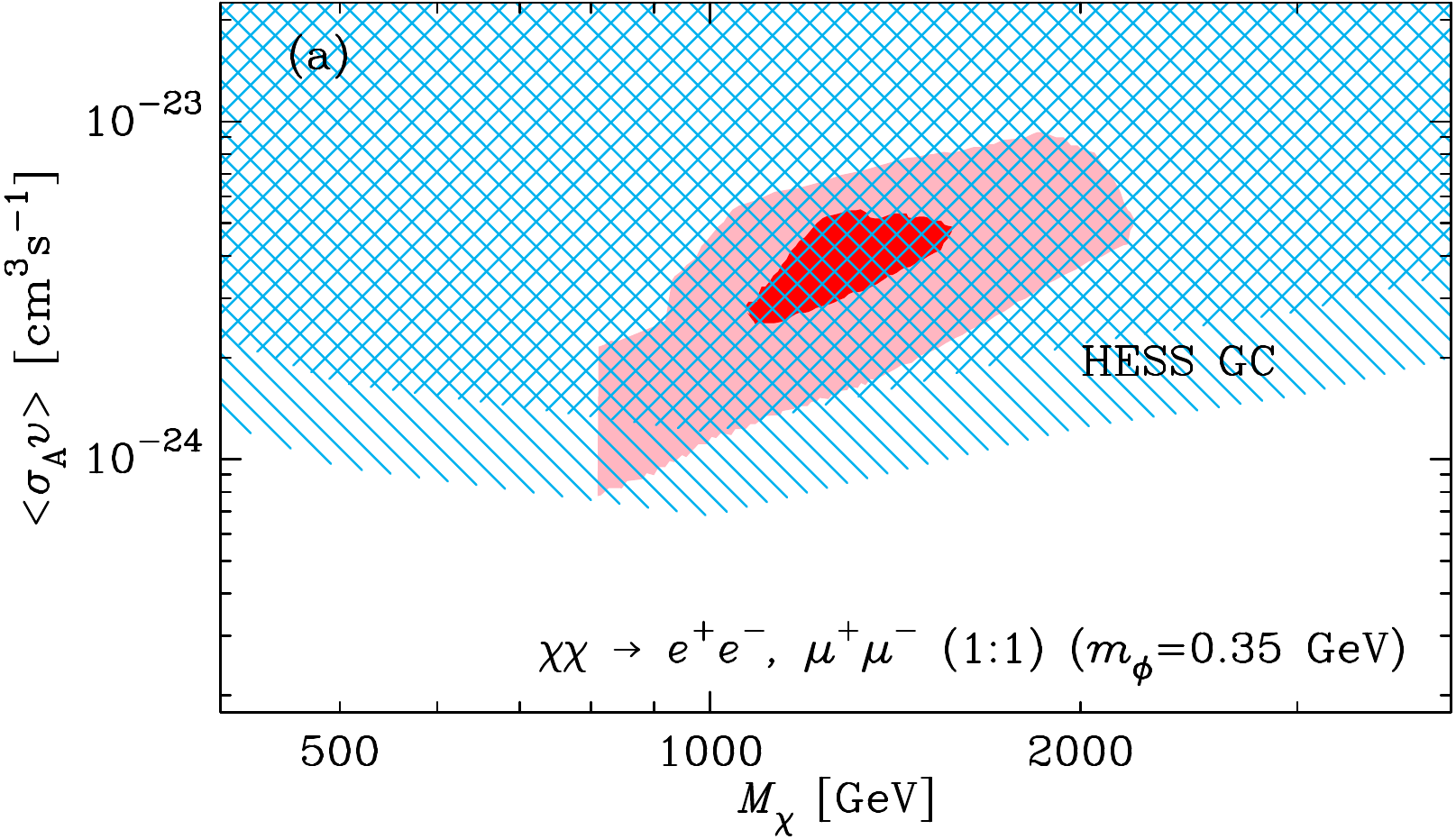} \\
\includegraphics[width=0.68\columnwidth]{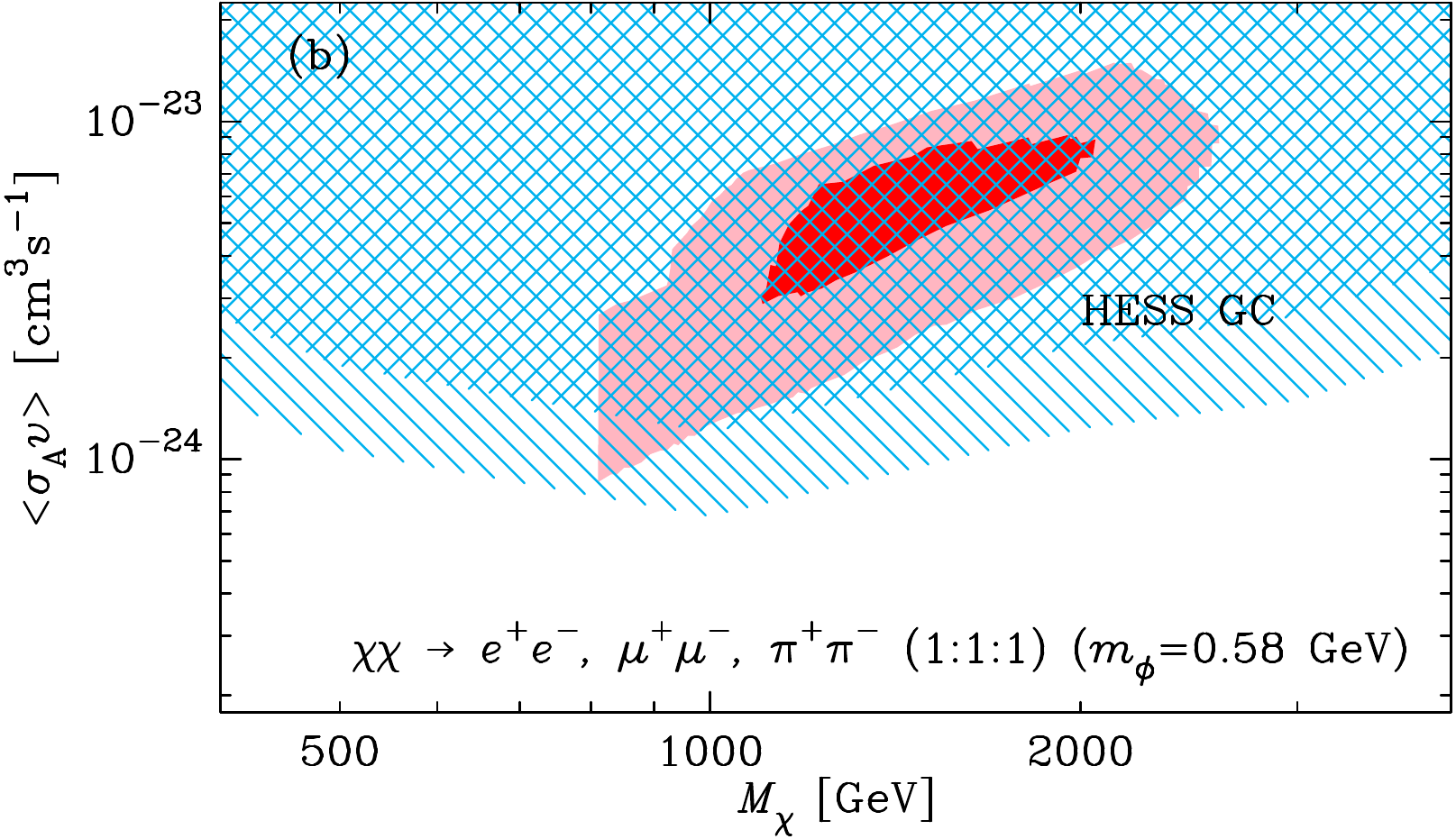} \\
\includegraphics[width=0.68\columnwidth]{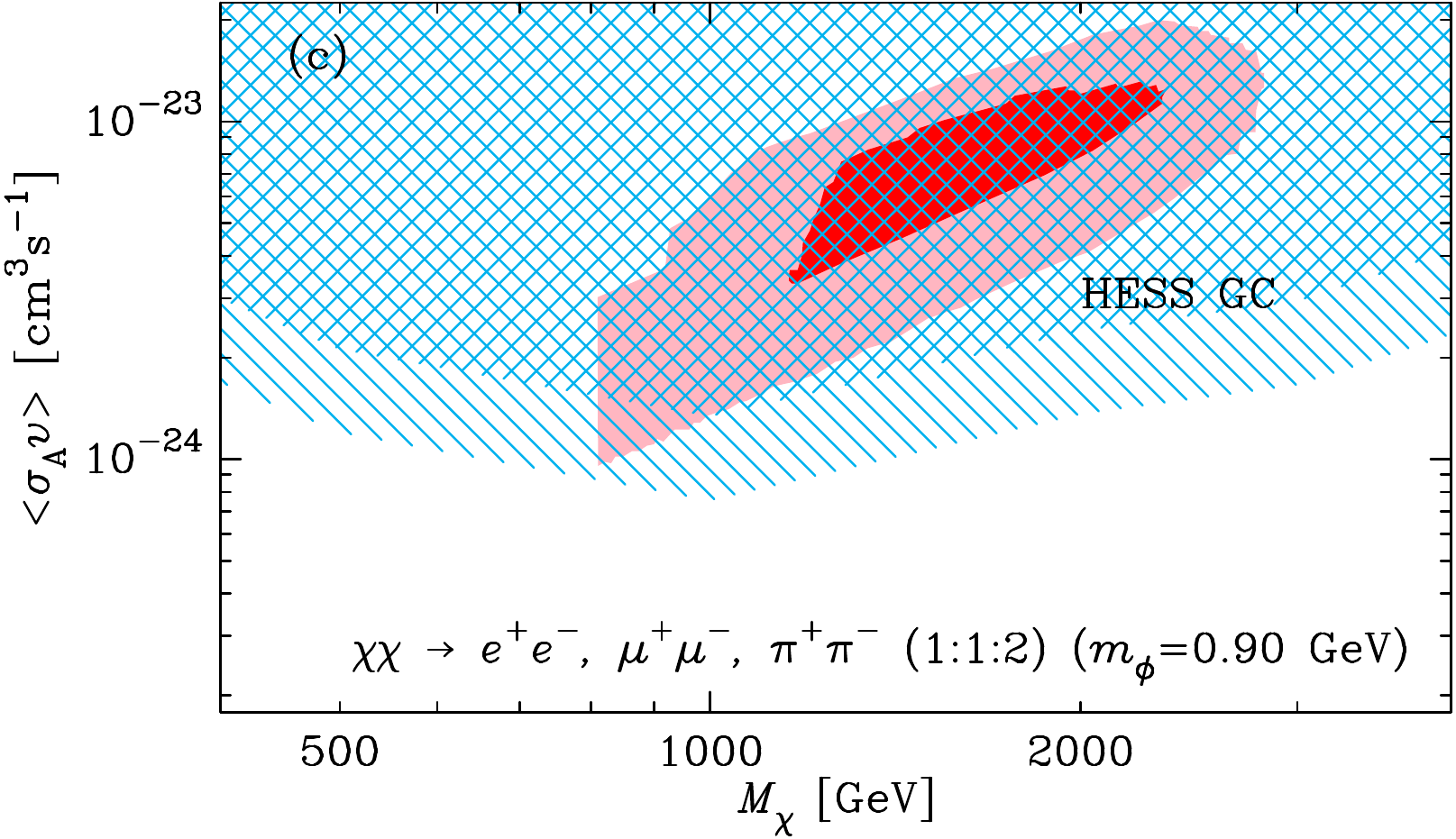}
\end{array}$
\end{center}
\caption{\small Shown are the constraints on dark matter in three XDM
  annihilation cases: (a) annihilation into $50\%$ $e^+e^-$ and
  $50\%$ $\mu^+\mu^-$ via two intermediate 0.35 GeV gauge bosons
  $\phi$; (b) annihilation into $33\%$ $e^+e^-$, $33\%$ $\mu^+\mu^-$,
  and $33\%$ $\pi^+\pi^-$ via two intermediate 0.58 GeV gauge bosons $\phi$; (c) annihilation into $25\%$ $e^+e^-$, $25\%$
  $\mu^+\mu^-$, and $50\%$ $\pi^+\pi^-$ via two intermediate 0.90 GeV
  gauge bosons $\phi$. The regions are labeled according to their
  constraining observations as described in the text: ``HESS GC'' are
  the $95\%$ CL limits from the HESS analysis of the GC. The double
  hatched region is constrained for both the Einasto and NFW halo
  models, and the single hatched region is constrained for only the
  Einasto halo model. The light pink and dark red shaded regions are
  the $68\%$ and $95\%$ CL regions consistent with the XDM dark matter
  interpretation of a combination of the PAMELA signal and the
  Fermi-LAT $e^+e^-$ feature from
  ref.~\cite{Finkbeiner:2010sm}.\label{XDM}}
\end{figure*}

In figure~\ref{XDM} three benchmark XDM models which are consistent
with a combination of the PAMELA signal and the Fermi-LAT feature are
shown, with red representing the $68\%$ CL region and light pink
representing the $95\%$ CL region. Figure~\ref{XDM}(a) contains the
regions for annihilations which go $50\%$ into $e^+e^-$ and $50\%$
into $\mu^+\mu^-$ through two intermediate gauge bosons of mass
$m_\phi=0.35\rm\ GeV$; figure~\ref{XDM}(b) contains the regions for
annihilations which go $33\%$ into $e^+e^-$, $33\%$ into $\mu^+\mu^-$,
and $33\%$ into $\pi^+\pi^-$ through two intermediate gauge bosons of mass
$m_\phi=0.58\rm\ GeV$; and figure~\ref{XDM}(c) contains the regions
for annihilations which go $25\%$ into $e^+e^-$, $25\%$ into
$\mu^+\mu^-$, and $50\%$ into $\pi^+\pi^-$ through two intermediate
gauge bosons of mass $m_\phi=0.90\rm\ GeV$.

When comparing to signal-fit regions from other work, a scaling of
$\rho_0^2/\rho_{\odot}^2$ has been done to normalize that work's local
dark matter density $\rho_0$ to the one adopted here. In the
literature, signals for the PAMELA and Fermi-LAT excesses are ascribed
boost factors for dark matter annihilation that can include both
astrophysical substructure boosts and particle physics boosts, as well
as the enhancement of latter due to the former. The local boost can be
separated as $B_{\rm local}=B_{s}B_{p}$ into the substructure boost
$B_{s}$ and the particle boost $B_{p}$ (when ignoring the enhancement
of the latter to to the former). To be clear and conservative, we
employ a relatively strong local substructure boost, $B_{s}=1.57$,
using that expected from unresolved substructure calibrated to the Via
Lactea II simulations~\cite{Kamionkowski:2010mi,Abazajian:2010zb}. We
incorporate the fit regions from ref.~\cite{Bergstrom:2009fa} such
that
\begin{equation}
E_F=\left(\frac{\rho_{\odot}}{0.3\rm \ GeV\ cm^{-3}}\right)^2B_{s}B_{p}\enspace.
\end{equation}
Both $B_{\rm local}$ and $E_F$ designate the scaling factor from
$\langle\sigma_{\rm A}v\rangle=3\times10^{-26}\rm\ cm^3\ s^{-1}$, as
shown in our figures.

Relative boosts between the GC and local effects can alter the
relative constraints between the annihilation in the GC versus local
cosmic ray signals in Sommerfeld-enhanced annihilation models, where
Sommerfeld-enhancement is greater in low-velocity substructures.  As
discussed in ref.~\cite{Slatyer:2011kg}, the boost toward the GC
relative to the local is
\begin{equation}
\frac{B_{\rm GC}}{B_{\rm local}} = \frac{S_{v(r=0)}/S_{v\sim
      150\rm\ km/s}}{1+(S_{v\rightarrow 0}/S_{v\sim
      150\rm\ km/s})\Delta(8.5\rm\ kpc)},
\label{gcvslocal}
\end{equation}
where $S$ is the Sommerfeld-enhancement in a given model at a distance
$r$ from the GC, for a given velocity dispersion of the dark matter
$v$. Here, $\Delta \equiv 1-B_s$, which we take to be
$\Delta(8.5{\rm\ kpc} = 0.57$.

For specificity, we adopt $m_\phi = 0.25\rm\ GeV$, which corresponds
to $S_{v\rightarrow 0}/S_{v\sim 150\rm\ km/s} \approx 5$, from figure
1 of ref.~\cite{Slatyer:2011kg}.  This model is a $4e$ channel case
shown in figure~\ref{4leptons}(a). The purple rectangle shows the
range of annihilation cross sections for a 1.2 TeV Sommerfeld-enhanced
scenario consistent with the thermal relic density, the CMB,
self-interaction bounds, and naturalness, for $\Delta$ from $10^{-4}$
to unity, corresponding to $B_{\rm local}$ from 30 to
300~\cite{Slatyer:2011kg}.  The scaling of the signal toward the GC
relative to the local boost is set explicitly by the local
substructure boost $\Delta$, and the value of $S_{v(r=0)}$, which is
determined by the velocity dispersion of dark matter in the
$\sim$450~pc of the HESS observation of the GC (and {\em not} exactly
at $r=0$).  It has been shown that the dark matter velocity rapidly
decreases toward the GC in Milky Way scale
halos~\cite{Navarro:2003ew}, and the limit $v\rightarrow 0$ is
potentially appropriate for $S_{v(r=0)}$, which we adopt in one case
of our limits.  In this GC low velocity limit case, one can take the
substructure dominant versus subdominant cases:
\begin{equation}
\frac{B_{\rm GC}({v\rightarrow 0})}{B_{\rm local}} \approx 
\begin{cases}
5\ \qquad (\Delta = 10^{-4}), \\
0.83 \quad (\Delta = 1) .
\end{cases}
\end{equation}
Baryonic effects have been found to enhance the velocity of the dark
matter toward the GC to make it comparable or greater than to that at
the solar distance~\cite{BoylanKolchin:2005wg,Pedrosa:2009rw}, such
that $S_{v(r=0)}\approx S_{v\sim 150\rm\ km/s}$ \cite{Cholis:2009va},
and in that scenario the substructure limiting cases are:
\begin{equation}
\frac{B_{\rm GC}({v\sim 150\rm\ km\ s^{-1}})}{B_{\rm local}} \approx
\begin{cases}
1\ \qquad (\Delta = 10^{-4}), \\
0.17 \quad (\Delta = 1) .
\end{cases}
\end{equation}
We designate these ranges of constraints as bars and arrows for the
corresponding four cases in figure \ref{4leptons}(a), with the
$v\rightarrow 0$ GC in green and $v\sim\rm 150\ km\ s^{-1}$ GC in dark
blue.  

Ref.~\cite{Slatyer:2011kg} claims that
$\Delta(8.5{\rm\ kpc})/\Delta(r=0) \sim 20$ is relevant for the
comparative constraints between the GC and the local boost, which
strictly is only the case for strong local substructure domination and
high velocities in the GC. More generally, it is eq.~\eqref{gcvslocal}
and the local substructure boost that sets the scaling, and is what we
adopt.  Note that any substructure in the dark matter of the GC region
would enhance the constraints here. We ignore this enhancement to be
conservative, i.e., we only include the annihilation from the smooth
component in the GC region.

\subsection{Discussion}
For the $b\bar{b}$ annihilation channel (figure~\ref{hadronic}(a)),
the HESS GC constraints are stronger than the constraints from the
Fermi-LAT analysis of dwarf spheroidal galaxies for $M_\chi\gtrsim
900\rm\ GeV$ when adopting a non-cored Einasto or NFW dark matter
profile. The limits on this channel for an Einasto dark matter profile
are within an order of magnitude of the thermal cross-section for
$2{\rm\ TeV}\lesssim M_\chi\lesssim 5\rm\ TeV$. Similarly, the
$W^+W^-$ annihilation channel (figure~\ref{hadronic}(c)) has stronger
constraints than the Fermi-LAT dwarfs for $M_\chi\gtrsim 800\rm\ GeV$
and limits $\langle\sigma_{\rm A}v\rangle\leq 3\times10^{-25}\rm\
cm^3\ s^{-1}$ for $1{\rm\ TeV}<M_\chi< 6\rm\ TeV$. The $t\bar{t}$
annihilation channel (figure~\ref{hadronic}(b)) has somewhat weaker
constraints, limiting $\langle\sigma_{\rm A}v\rangle\leq
5\times10^{-25}\rm\ cm^3\ s^{-1}$ for $2{\rm\ TeV}<M_\chi< 10\rm\
TeV$.

The light pink PAMELA excess region of ref.~\cite{Meade:2009iu} and
the light green PAMELA excess region of ref.~\cite{Bergstrom:2009fa}
are both excluded above $M_\chi\sim 400\rm\ GeV$ by the HESS GC data
for the $\mu^+\mu^-$ annihilation channel
(figure~\ref{leptonic}(a)) when adopting a non-cored
Einasto or NFW halo model. Also for the $\mu^+\mu^-$ channel, the red
Fermi-LAT feature region of ref.~\cite{Meade:2009iu} and the dark
green Fermi-LAT feature region of ref.~\cite{Bergstrom:2009fa} are
excluded by the HESS GC data when adopting either halo model.

In the $\tau^+\tau^-$ annihilation channel (figure~\ref{hadronic}(b)),
the HESS GC observation excludes a cross-section of
$\langle\sigma_{\rm A}v\rangle>4\times10^{-26}\rm\ cm^3\ s^{-1}$ for a
dark matter mass $M_\chi\approx 1\rm\ TeV$, when adopting an Einasto
profile, within a factor of $\sim$2 of the thermal cross-section.  The
HESS GC excludes the light pink PAMELA excess region above
$M_\chi\approx 400\rm\ GeV$ and the Fermi-LAT dwarf analysis excludes
the $\tau^+\tau^-$ channel below $M_\chi\approx 400\rm\ GeV$, so this
model for the PAMELA excess has been ruled out at all dark matter
masses when adopting an NFW profile. The red Fermi-LAT feature region
is also excluded, at greater than 99.9999\% CL, in the $\tau^+\tau^-$
channel for both the Einasto and NFW dark matter profiles, consistent
with previous
results~\cite{Bertone:2008xr,Meade:2009iu,Abazajian:2010sq}.

For the $4e$ annihilation channel (figure~\ref{4leptons}(a)), the HESS
GC observation constrains the light pink PAMELA excess region of
ref.~\cite{Meade:2009iu} above $M_\chi\approx 600\rm\ GeV$ for the Einasto
halo model, and is constrained above $M_\chi\approx 500\rm\ GeV$ for the
NFW halo model. Similarly, the $4\mu$ annihilation channel
(figure~\ref{4leptons}(b)) has both the light pink region and the
light green PAMELA excess region of ref.~\cite{Bergstrom:2009fa}
constrained above $M_\chi\approx 1.5\rm\ TeV$ for the Einasto halo model, and
above $M_\chi\approx 3\rm\ TeV$ for the NFW halo model. 

In the $4e$ channel (figure~\ref{4leptons}(a)), the purple rectangle
shows the range of boost factors for a 1.2 TeV Sommerfeld-enhanced
model consistent with the thermal relic density, the CMB,
self-interaction bounds, and naturalness, for $\Delta$ conservatively
from $10^{-4}$ to unity~\cite{Slatyer:2011kg}. Note that the gamma-ray
annihilation constraints from the HESS GC on the annihilation
cross-section and boost factor, when adopting an NFW or Einasto
profile, is often stronger than those from the thermal relic density,
the CMB, self-interaction bounds, and naturalness considerations.
Specifically, the limits to boosts toward the GC are $B_{\rm GC}
\lesssim 27$, which for the comparable central velocity case excludes
models with local substructure $\Delta \lesssim 0.025\ (0.17)$ for
$B_{\rm local} = 30\ (50)$.  Our canonical case for $\Delta \approx
0.57$, with GC dark matter velocities comparable to that locally, is
unconstrained at 95\% CL.  Note that any substructure within the inner
degree of the HESS GC observation would further enhance these limits
due to the corresponding Sommerfeld-enhancement saturation in the GC
substructure. The substructure boost in the inner degree has been seen
in simulations at the level of $\Delta(r\sim 0.15{\rm\ kpc}) \sim
4\times 10^{-2}\ \text{to}\ 4\times 10^{-4}$~\cite{Kuhlen:2008aw}.

The XDM annihilation models: ${\rm Br}(e^+e^-)={\rm
  Br}(\mu^+\mu^-)=0.5$, $m_\phi=0.35\rm\ GeV$ (figure~\ref{XDM}(a));
${\rm Br}(e^+e^-)={\rm Br}(\mu^+\mu^-)={\rm Br}(\pi^+\pi^-)=0.33$,
$m_\phi=0.58\rm\ GeV$ (figure~\ref{XDM}(b)) ; and, ${\rm
  Br}(e^+e^-)={\rm Br}(\mu^+\mu^-)=0.25$,${\rm Br}(\pi^+\pi^-)=0.5$,
$m_\phi=0.90\rm\ GeV$ (figure~\ref{XDM}(c)) are excluded at greater
than $95\%$ CL when adopting an Einasto halo profile. However, when
adopting an NFW halo profile, these XDM models have a region from
$0.8{\rm\ TeV}\lesssim M_\chi\lesssim 1\rm\ TeV$ below
$\langle\sigma_{\rm A }v\rangle\sim 1.8\times10^{-24}\rm\ cm^3\
s^{-1}$ which remains consistent at $95\%$ CL.  
\section{Conclusions}
The HESS telescope's observations toward the Galactic center present
the strongest constraints on WIMP dark matter annihilation into
Standard Model particles for $M_\chi \gtrsim 900\rm\ GeV$, given a
non-adiabatically-contracted NFW or Einasto profile for the Milky Way
dark matter profile.  As discussed in the introduction, the HESS GC
observation is not sensitive to dark matter annihilation for large
constant-density cored dark matter profiles.  If such profiles
are established to be valid in the Milky Way, the HESS GC observation
provides no empirical constraint in these cases given the background
subtraction method.  If adiabatic contraction and steepening of the
dark matter density profile is well established, then the constraints
could become stronger.  We find that constraints on annihilation into
$b\bar b$ final states comes within an order of magnitude of the
canonical thermal cross section, while $t\bar t$ is within a factor of
$\sim$20. The $\tau^+\tau^-$ channel is within a factor of $\sim$2 of
the thermal cross-section.  This bodes well for the future
\v{C}erenkov Telescope Array's potential impact at constraining
canonical thermal WIMP dark matter this mass
scale~\cite{Consortium:2010bc}.

We also examine constraints on Sommerfeld-enhanced dark matter
annihilation models which produce the PAMELA positron excess and
Fermi-LAT $e^+e^-$ spectral feature, including XDM.  The models with
pure leptonic modes $\tau^+\tau^-$ and 4$e$ are excluded at greater
than 95\% CL by the HESS GC when adopting NFW or Einasto dark matter
halo profiles.  For other cases, (e.g. 4$\mu$, $2e2\mu$ , $e^+ e^-
\mu^+ \mu^- \pi^+ \pi^-$), the models are in tension with the HESS GC
observations, with portions of the signal 95\% CL parameter space are
excluded at the 95\% CL level.  Significantly, the exclusions
presented here from HESS GC gamma-ray observations on
Sommerfeld-enhanced upper boosts are more constraining, in many cases,
than prior constraints from diffuse
gamma-rays~\cite{Abazajian:2010zb,Zavala:2011tt}, relic density
considerations~\cite{Dent:2009bv,Zavala:2009mi,Feng:2009hw,Buckley:2009in,Feng:2010zp},
the
CMB~\cite{Padmanabhan:2005es,Zavala:2009mi,Slatyer:2009yq,Hisano:2011dc},
halo shapes~\cite{Feng:2009hw,Buckley:2009in}, and
naturalness~\cite{Finkbeiner:2010sm,Slatyer:2011kg}.  In the case
where the velocity dispersion of dark matter in the center of the
Galaxy is comparable to that locally, the supplemental
Sommerfeld-enhancement from local substructure and not at the GC
weakens the constraints in certain models.

Gamma-ray astronomy has produced the most stringent constraints on the
canonical thermal WIMP model's annihilation cross-section, with
Fermi-LAT's stacked observations of dwarf galaxies being the most
constraining at low masses, and HESS's observations of the Galactic
center being the most constraining at higher masses.  With further
observation and new technologies, the nature of dark matter may be
revealed by gamma-ray telescopes.

\begin{acknowledgments}
  We thank Prateek Agrawal, Mike Boylan-Kolchin, Oleg Gnedin, Manoj
  Kaplinghat, Andrey Kravtsov, Mike Kuhlen, Tracy Slatyer and Neal
  Weiner for useful discussions. KNA \& JPH are supported by NSF Grant
  07-57966 and NSF CAREER Award 09-55415.
\end{acknowledgments}

\bibliography{bibliography}
\end{document}